\documentclass[twocolumn,prl,amsmath,superscriptaddress,nofootinbib,amssymb]{revtex4}
\usepackage[usenames]{color}
\definecolor{simon}{cmyk}{1.0, 0.0, 1.0, 0.0} 

\definecolor{norm}{cmyk}{0,0,0,1}

\definecolor{luca}{cmyk}{0.0, 1.0, 1.0, 0.0} 

\usepackage{graphicx}
\usepackage{times}
\usepackage{dcolumn}

\begin{document}

\title {Quasiparticles and charge transfer at the two surfaces of the honeycomb iridate Na$_2$IrO$_3$}

\author{L. Moreschini}\email{lmoreschini@lbl.gov}
\affiliation{Advanced Light Source, Lawrence Berkeley National Laboratory, Berkeley, California 94720, USA}

\author{I. Lo Vecchio}
\affiliation{Materials Sciences Division, Lawrence Berkeley National Laboratory, Berkeley, California 94720, USA}

\author{N. P. Breznay}
\affiliation{Materials Sciences Division, Lawrence Berkeley National Laboratory, Berkeley, California 94720, USA}
\affiliation{Department of Physics, University of California Berkeley, Berkeley, California 94720, USA}

\author{S. Moser}
\affiliation{Advanced Light Source, Lawrence Berkeley National Laboratory, Berkeley, California 94720, USA}

\author{S. Ulstrup}
\affiliation{Advanced Light Source, Lawrence Berkeley National Laboratory, Berkeley, California 94720, USA}

\author{R. Koch}
\affiliation{Advanced Light Source, Lawrence Berkeley National Laboratory, Berkeley, California 94720, USA}

\author{J. Wirjo}
\affiliation{Materials Sciences Division, Lawrence Berkeley National Laboratory, Berkeley, California 94720, USA}
\affiliation{Department of Physics, University of California Berkeley, Berkeley, California 94720, USA}

\author{C. Jozwiak}
\affiliation{Advanced Light Source, Lawrence Berkeley National Laboratory, Berkeley, California 94720, USA}

\author{K. S. Kim}\affiliation{Department of Physics, Yonsei University, Seoul 03722, Korea}

\author{E. Rotenberg}
\affiliation{Advanced Light Source, Lawrence Berkeley National Laboratory, Berkeley, California 94720, USA}

\author{A. Bostwick}
\affiliation{Advanced Light Source, Lawrence Berkeley National Laboratory, Berkeley, California 94720, USA}

\author{J. G. Analytis}
\affiliation{Materials Sciences Division, Lawrence Berkeley National Laboratory, Berkeley, California 94720, USA}
\affiliation{Department of Physics, University of California Berkeley, Berkeley, California 94720, USA}

\author{A. Lanzara}
\affiliation{Materials Sciences Division, Lawrence Berkeley National Laboratory, Berkeley, California 94720, USA}
\affiliation{Department of Physics, University of California Berkeley, Berkeley, California 94720, USA}

\begin{abstract}

Direct experimental investigations of the low energy electronic structure of the Na$_2$IrO$_3$ iridate insulator are sparse and draw two conflicting pictures. One relies on flat bands and a clear gap, the other involves dispersive states approaching the Fermi level, pointing to surface metallicity. Here, by a combination of angle resolved photoemission, photoemission electron microscopy and x-ray absorption, we show that the correct picture is more complex and involves an anomalous band, arising from charge transfer from Na atoms to Ir-derived states. Bulk quasiparticles do exist, but in one of the two possible surface terminations the charge transfer is smaller and they remain elusive.
\end{abstract}
\maketitle

The honeycomb iridate Na$_2$IrO$_3$ represents an ideal example of a 5$d^5$ system with complete removal of the orbital degeneracy by the spin-orbit interaction \cite{balents}, and for this reason has been object of considerable attention in recent years. Magnetically, it has been proposed to realize the Kitaev model due to the hexagonal symmetry of the $ab$ planes. The ground state was instead shown to have a zigzag antiferromagnetic (AF) order \cite{liu,ye,choi}, accounted for by direct 5$d$-5$d$ overlap \cite{chaloupka2010}, next nearest-neighbor coupling \cite{kimchi,foyevtsova,singh2012,choi,kargarian}, or possibly inter-orbital hopping \cite{chaloupka2013}. Electronically, theoretical calculations have tentatively categorized this compound as a topological insulator \cite{shitade,yu}. Transport and optical measurements suggest rather a Mott insulator picture \cite{singh2010,singh2012,sohn,gedik}, and an ongoing debate exists on whether spin-orbit coupling plays a decisive role \cite{chaloupka2010,yunoki,gretarssonprl} or a collaborative one \cite{mazin2012,mazin2013,foyevtsova,sohn} in opening the gap. 

The latter point essentially comes down to whether a relativistic approach with two effective J$_\mathrm{eff}$=1/2 and J$_\mathrm{eff}$=3/2 levels, as generally accepted in the description of the Ruddlesden-Popper series iridates \cite{bjprl}, holds for this honeycomb lattice, or if spin-orbit coupling only ``assists'' a band gap. While experiments targeting the magnetic order have been numerous and exhaustive \cite{ye,liu,singh2010,gretarssonprb,chun}, measurements of the electronic structure have not been as successful. Unlike in perovskite iridates, which with a more symmetric crystal structure represent an ideal playground for angle resolved photoemission (ARPES) \cite{king,wang,simon,mio,delatorre1,yeongkwan,he}, photoemission data for Na$_2$IrO$_3$ are limited to two instances \cite{comin,alidoust}, owing to the difficulty of obtaining sufficiently large cleaves.

In one case, only remarkably flat bands with $<$100 meV bandwidth were observed, and no quasiparticles. The Fermi level was pinned at the top of a valence band with a density of states (DOS) reminiscent of a pseudogap \cite{comin}. In the other case, bands dispersing over more than 1 eV, more compatible with typical $5d$ bandwidths, were found, and more importantly a weak intensity in the vicinity of the Fermi level, which suggested surface metallicity \cite{alidoust}. In this Letter we use spatially-resolved ARPES to determine the electronic structure of Na$_2$IrO$_3$ and its dependence upon the surface termination.  We find that a clear quasiparticle at the Fermi level can be measured owing to charge transfer from the Na atoms, but only for Na-terminated surfaces. For O-terminated surfaces, the charge transfer between Na and Ir is reduced, leading to a strong suppression of the quasiparticle and a large gap.

\begin{figure}[t]
\includegraphics[width=8.5cm]{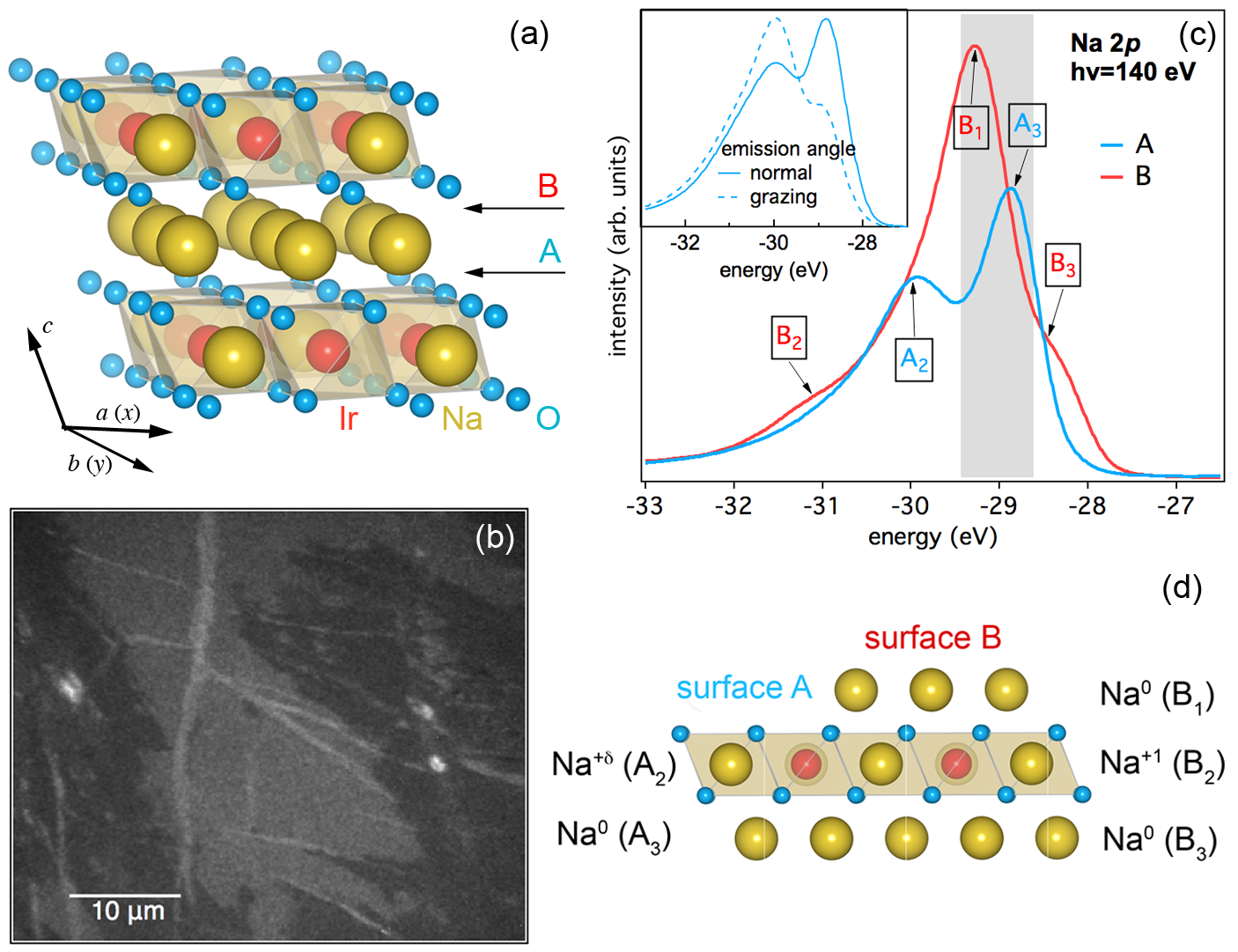}
\caption{\label{fig1} (a) crystal structure of Na$_2$IrO$_3$. The A and B arrows indicate the two possible cleavage points for the crystal; (b) a PEEM image measured on a sample cleaved in UHV. The energy window of the photoelectrons selected by the aperture is indicated by the shaded area in (c); (c) Na 2$p$ spectra measured on the two surface terminations indicated in (a). The inset shows, for surface A, two spectra taken at different emission angles; (d) Stacking of the top layers for both A and B. Next to each layer we indicate the valence of the Na atoms referred to a relative scale (see text) and the label of the corresponding peak in (c).}
\end{figure}

In Na$_2$IrO$_3$ the characteristic building blocks of iridates, the IrO$_6$ octahedra, are edge-sharing and form a layered stacking alternating with pure Na layers. The crystal structure, with $C2/m$ space group, is less symmetric than in cubic perovskites. As shown in Fig.~\ref{fig1}(a), it presents a monoclinic $c$ axis tilted by $\sim109^\circ$ \cite{choi}. The interlayer hopping terms are negligible, as we will discuss later, and therefore we will refer throughout this paper to the hexagonal surface Brillouin zone (BZ) of the $ab$ honeycomb lattice, with lattice constant $a=5.427\,\mathrm{\AA}$. The surface magnetic unit cell, arising from the zigzag AF order, is rectangular and twice as small in $k$ space, but as we will show the nonmagnetic BZ is the logical choice, since the electron periodicity is clearly hexagonal.

Inspecting the crystal structure of Fig.~\ref{fig1}(a), it appears evident that two cleavage planes are geometrically equivalent and therefore equally probable. They are indicated as A and B in the figure and throughout the rest of this Letter, and correspond to an exposed Na$_{0.5}$IrO$_3$ and Na$_{1.5}$ surface, respectively. In a standard band insulator/semiconductor, one should expect from ARPES to observe the same bulk band structure (possibly with a different band bending) and different surface electronic bands depending on the termination \cite{bitei}. For this Mott insulator we will draw instead a more intricate picture.

The presence of two different surfaces can be easily verified by x-ray photoemission spectroscopy (XPS) of the Na 2$p$ levels, as discussed in the following. Therefore these can be mapped in real space with a photoemission electron microscope (PEEM). The image in Fig.~\ref{fig1}(b) has been recorded in a dark field mode by accepting through the lens system only the photoelectrons in the energy range indicated by the shaded area in Fig.~\ref{fig1}(c), so that the intensity difference in XPS provides the color contrast in the dark field image. We found separate regions of a typical size of 10 to 40 $\mu$m, but small cracks and spurious areas are almost always present. The core level data in Fig.~\ref{fig1}(c) and the ARPES data shown next were obtained in a $\mu$ARPES endstation with a spot size of $\sim$10-20$\,\mu$m, sufficient to separate the signal from the two domains. At first we discuss the core level intensities in the context of a structural model, and then we explain the binding energies and their correlation to the electronic structure. These two aspects give a self-consistent qualitative understanding of all of the valence band data discussed later.

Experimentally we observe for cleaving plane A two components, as shown in Fig.~\ref{fig1}(c). These two peaks, A$_2$ and A$_3$, are assigned to the outermost Na$_{0.5}$IrO$_3$ layer of the octahedral-terminated surface [Fig.~\ref{fig1}(d)], and the next deeper, pure Na$_{1.5}$ layer, respectively. The higher intensity of A$_3$ is accounted for by the number of Na atoms, which is three times larger in A$_3$ than A$_2$, but this intensity ratio is reduced from 3 to $\sim$1.5 by the attenuation of the A$_3$ signal due to its depth. This assignment  is confirmed by XPS at grazing emission, in which A$_3$ is further suppressed [inset of Fig.~\ref{fig1}(c)]. For the B surface two weak photoemission peaks (B$_2$, B$_3$) are observed with a very similar branching ratio as A$_2$:A$_3$, and are ascribed to Na atoms in the same respective layers, now buried by an additional Na$_{1.5}$ layer, with corresponding peak B$_1$ largely dominant.  The overall weaker intensities of (B$_2$, B$_3$) compared to (A$_2$, A$_3$) are explained simply by attenuation due to this new outer layer. 

The binding energies of all the peaks can be understood by simple considerations on the initial and final states of photoemission from ionic insulators \cite{eli}, where the valence of the different Na atoms are indicated in the model of Fig.~\ref{fig1}(d). Na is least oxidized in the Na$_{1.5}$ layers and most oxidized in the Na$_{0.5}$IrO$_3$ layers, where it donates charge to the IrO$_6$ octahedra. For the sake of argument we call these valences ``0'' and ``+1'', respectively, although these should be understood on a relative scale.  

For the core level peaks B$_3$, A$_3$ and B$_1$ (listed in order of increasing binding energy) of the structurally and chemically equivalent Na$_{1.5}$ layers, the variation of position corresponds to the progressively weaker screening response to the core hole as the surface is approached \cite{eli}. The remaining peaks B$_2$ and A$_2$, representative of the Na$_{0.5}$IrO$_3$ layers, are shifted to higher binding energy due to charge transfer from Na to the IrO$_6$ octahedra. The decrease in binding energy of A$_2$ compared to B$_2$ is due to the lower valence of the former ($+\delta$ with $\delta$$<$1) compared to the latter (+1) due to the altered environment of the surface octahedra which causes a smaller charge transfer. We will show that this smaller charge transfer within the Na$_{0.5}$IrO$_3$ layer at the A surface compared to B (and to the bulk) has profound consequences on the bands observed by ARPES on this compound, since it hides the quasiparticles of the bulk electronic structure. 

\begin{figure*}[t!]
\includegraphics[width=18cm]{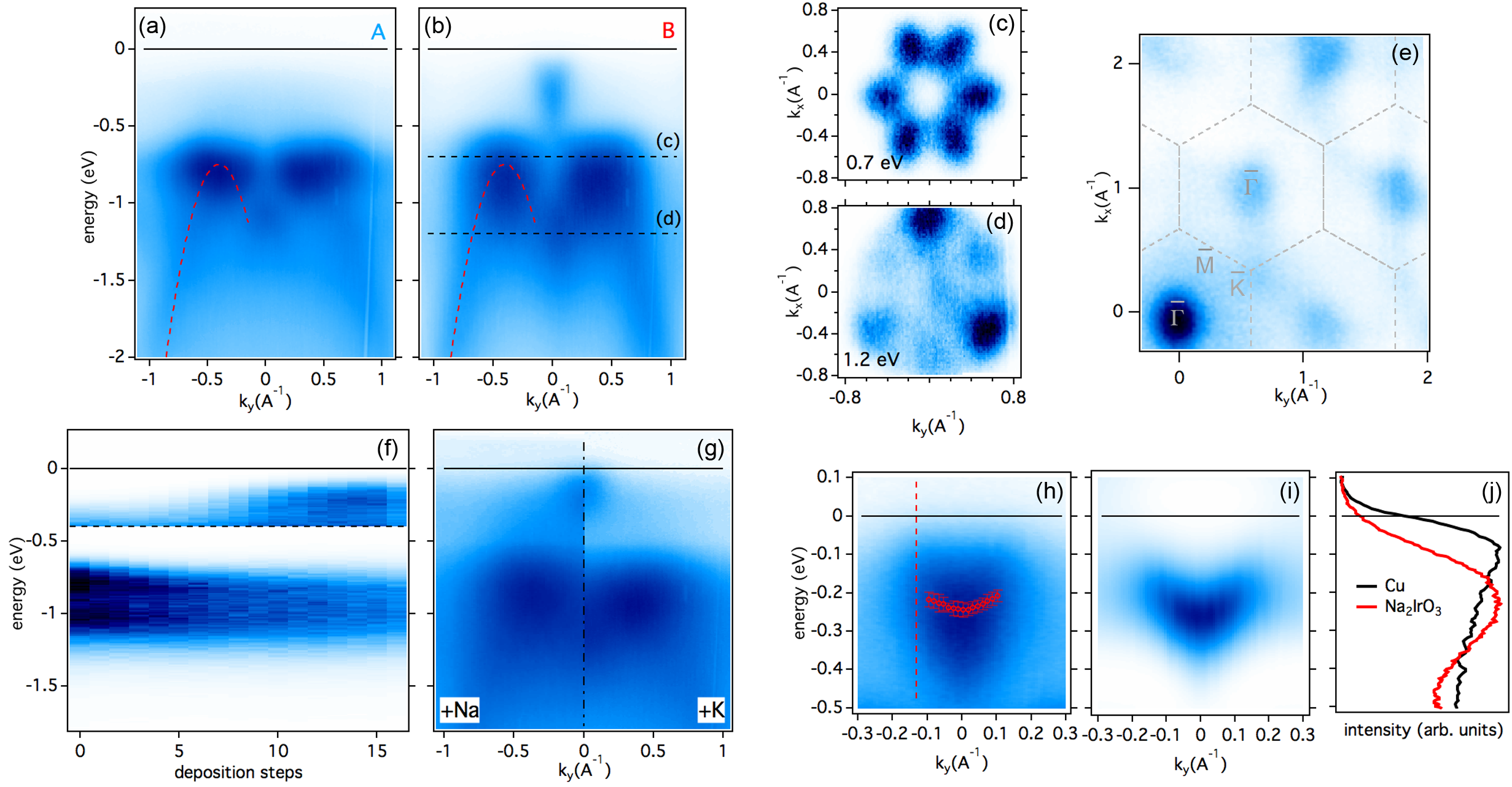}
\caption{\label{fig2}(a,b) valence band dispersion along $k_y$, measured at $\overline\Gamma$ for the A and B terminations. The dashed curves are guides to the eye for the hole-like dispersive states; (c,d) constant energy maps in the vicinity of $\overline\Gamma$ at the two energies indicated by the dashed lines in (a); (e) wide range constant energy map measured on surface B at 60 meV binding energy. The periodicity is the one of the nonmagnetic surface BZ with $\overline{\Gamma\mathrm{M}}=\pi/a\simeq0.58\,\mathrm{\AA}^{-1}$ and $\overline{\Gamma\mathrm{K}}\simeq0.67\,\mathrm{\AA}^{-1}$; (f) evolution of the EDC intensity at $\overline\Gamma$ with Na deposition on surface A. Two different color scales have been used above and below the horizontal dashed line to enhance the weak low-energy state with respect to the other valence states; (g) The band structure at $\overline\Gamma$ after Na and K deposition is shown on the left and right side of the image, respectively; (h) close up of the surface state in surface B. The markers indicate the peak positions as extracted by the EDC fits and the error bars define a 98\% confidence level; (i) image obtained applying the curvature method to (h); (j) The EDC at $k_y$ indicated by the dashed line in (h) is compared to the Fermi level measured on the Cu sample holder. The data in (h-j) were measured at $T\simeq200$K. Note that in all the ARPES figures $k_y$, with $y$ defined as in Fig.~\ref{fig1}(a), is the horizontal axis}
\end{figure*}

Figure \ref{fig2}(a,b) shows the band structure measured at $\overline{\Gamma}$ with the sample oriented along $\overline{\Gamma\mathrm{K}}$ for both A and B terminations. All the spectral features, similarly to the layered iridate perovskites \cite{king,mio}, proved to be hardly dispersive along the $k_z$ direction normal to the surface, hence our choice to discuss the data only in terms of the surface BZ. We set the photon energy between 80 and 90 eV in all the measurements to maximize the intensity of the valence states. Likewise, no obvious change was observed as a function of temperature down to $\sim$100K, where the sample started to show some charging. 

The valence band presents clearly dispersive hole-like states with maximum at $\sim$0.7 eV. These were seen in Ref.~\onlinecite{alidoust} but missed in Ref.~\onlinecite{comin} with a He lamp as excitation source, due to a small cross section. The flat bands seen in Ref.~\onlinecite{comin} are instead generally weak at higher photon energy \cite{alidoust}. Here, they gain a strong intensity at $\sim$0.7-0.9 eV, where they overlap with the maximum of the hole-like parabolae. A hint of the different orbital character of these two states -- the flat bands and the dispersive parabolae -- comes from the constant energy maps in Fig.~\ref{fig2}(c,d). The hexagonal contours are 30$^\circ$ rotated with respect to each other and in the higher energy one a three-fold modulation is clearly visible.

The most apparent difference between Fig.~\ref{fig2}(a) and Fig.~\ref{fig2}(b) is that the B surface hosts an additional electronic state with respect to the A surface. A wide range map at $\sim$60 meV below the Fermi level [Fig.~\ref{fig2}(e)] proves that this band is centered at $\overline{\Gamma}$ and no other potentially metallic states exist in other points of the reciprocal space. Some residual high intensity spots hint at a ($\sqrt{3}\times\sqrt{3})\mathrm{R}30^\circ$ reconstruction, as seen by scanning tunneling microscopy (STM) \cite{lupke}. Notice also that clearly the fine details in the 0.7-1.2 eV binding energy range are different between A and B, indicative of a different hybridization between the dispersive and the flat bands. 

We propose that the determining factor in the formation of the quasiparticles is the presence (B) or the absence (A) of the outermost  Na$_{1.5}$ layer. When present, this layer regenerates for the top Na$_{0.5}$IrO$_3$ layer the same coordination as in the bulk and the quasiparticles are observed. The Ir-O octahedra are embedded between two Na layers, as opposed to A where the absence of the Na overlayer appears to hinder the charge transfer from Na to Ir within the Na$_{0.5}$IrO$_3$ plane and therefore to hide the bulk low-energy state. A critical test of this hypothesis and of our structural model is to deposit Na atoms on the A termination and see if the electronic structure of B is recovered. 

Alkali atoms are frequently used with ARPES on metals or semiconductors to access more states in the conduction band or the value of the band gap, respectively. Here, certainly due to a different pinning of the Fermi level on the bare surface, the total energy shift upon Na deposition [see the slight movement of the energy distribution curve (EDC) peak at $\sim$0.8 eV in Fig.~\ref{fig2}(f)] is only $\sim$100 meV, but the most striking effect is the gradual appearance of the low energy band, as predicted above. Notice that the intensity does not cross the Fermi level from above, as in a rigid shift of the conduction band with the progressive addition of electrons, but forms instead a new electronic state. This effect is not element-specific either, and seems rather linked only to the $sp^1$ valence of the Na atoms. By physisorption of K, which is not present in the bulk material, exactly the same end point can be reached [see Fig.~\ref{fig2}(g)]. The evolution of the Na core levels upon Na and K deposition, as shown in the supplementary information \cite{supplinfo}, is fully consistent with our discussion of Fig.~\ref{fig1}.

Previous ARPES studies \cite{alidoust,comin} were conducted with larger probe sizes and therefore averaged signals from both A and B terminations. Their interpretation of the low energy states was constrained by a weaker signal to background ratio and some key details remained hidden. In Ref.~\onlinecite{alidoust} the dispersion was uncertain and was tentatively assigned to a hole-like band. No clear Fermi level crossing could be identified, in which case surface metallicity would arise only via thermally excited carriers. The DOS at the Fermi cut looked like a pseudogap leading edge, similarly to Ref.~\onlinecite{comin} which saw no in-gap states. This in turn led to a scenario where quasiparticles would be suppressed by the interference between the two sublattices of the honeycomb structure \cite{trousselet}. 

Figure~\ref{fig2}(h) shows a more detailed measurement of the region close to the Fermi level. In this image the dispersion is unambiguously electron-like. Notice that even at first sight it is incompatible with a quantum well state as observed, for instance, on the surface of Sr$_2$IrO$_4$ upon K deposition (supplementary information of Ref.~\onlinecite{yeongkwan}). The EDC fits superposed to the image plot, as well as the curvature method treatment shown in Fig.~\ref{fig2}(i) \cite{zhang}, give a positive effective mass of $\sim$$0.9\,m_{e}$. It would be tempting to extract from the band dispersion a 2D charge density ($\sim$$1.5\times10^{15}$ cm$^{-2}$), but this would imply a Fermi level crossing which in fact does not occur. 

In Fig.~\ref{fig2}(j) we plot the EDC at the $k$ vector of the supposed Fermi level crossing. Since the actual band dispersion cannot be followed reliably outside the range fitted in Fig.~\ref{fig2}(h), the EDC is integrated over a safe window of $0.08\,\mathrm{\AA}^{-1}$. Despite the relatively high ($\sim$200K) sample temperature, the slope of the leading edge cannot be explained with the natural broadening of the Fermi-Dirac function, as shown by the Fermi level measured on the Cu sample holder at the same temperature. Whether the lineshape should be better referred to as a gap or a pseudogap \cite{trousselet}, it cannot be explained by a surface state derived from the Na $sp$ electrons. 

\begin{figure}[t!]
\includegraphics[width=8.5cm]{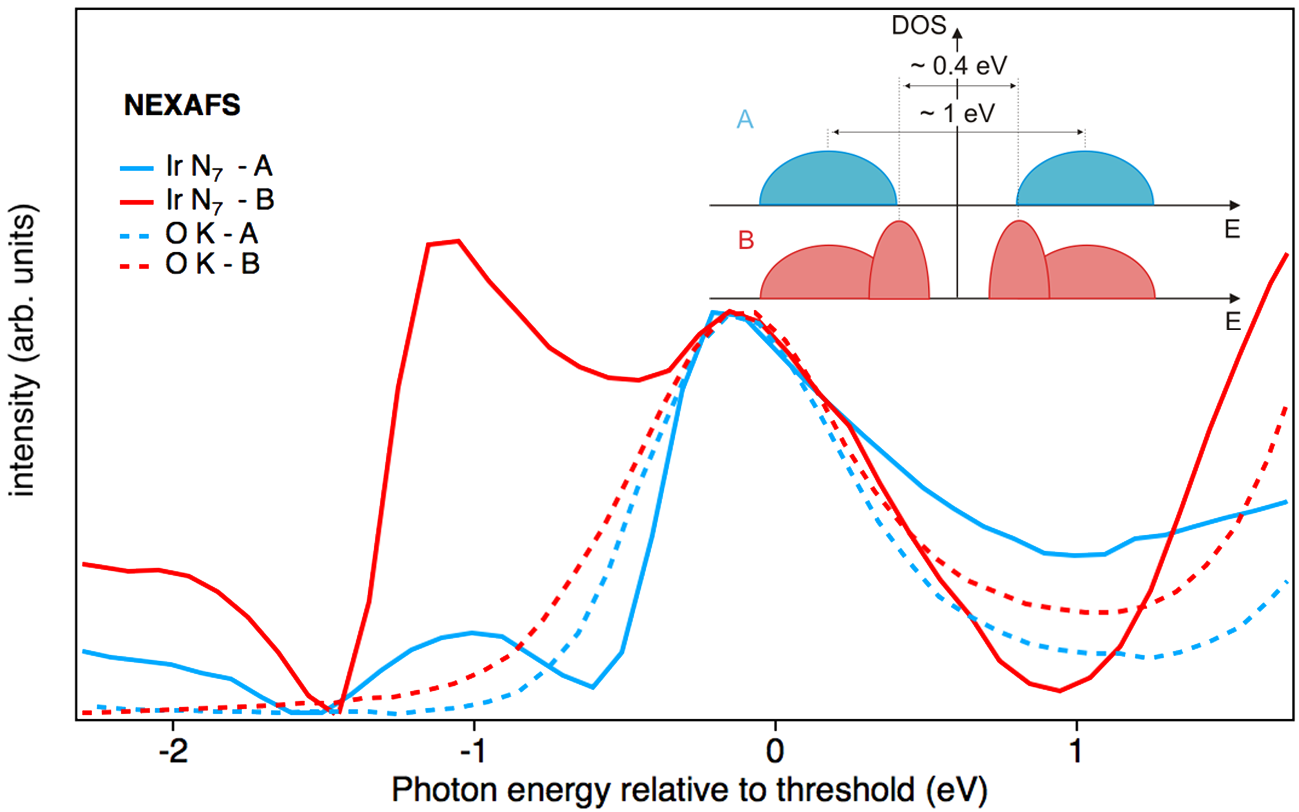}
\caption{\label{fig3}  XAS spectra at the N$_7$ (full line) and K (dashed line) edges measured on A and B are shown in blue and red, respectively. The photon energy on the horizontal axis has been offset by the binding energy of the final state core hole, namely 62.3 eV for Ir 4$f$$_{7/2}$ and 529.3 eV for O 1$s$; the inset shows a qualitative sketch of the DOS at the two surfaces after addition of charge to the nominal $d^5$ configuration, as inferred by ARPES and XAS. Note that the symmetry between occupied and unoccupied states is clearly an oversimplification.} 
\end{figure}

An elegant way of exploring the character of the low-energy state is by employing the chemical sensitivity of x-ray absorption (XAS). Although the strength of the $l\rightarrow l-1$ excitation channel is notoriously low at these energies \cite{minar}, Ir allows for the ``unusual'' transition $4f\rightarrow5d$ which provides a reasonably good intensity. Therefore we used the N$_7$ edge to resonantly probe the portion of the conduction band with Ir character. The data for both the A and B surfaces, obtained by a standard total electron yield measurement, are shown in Fig.~\ref{fig3}, together with those measured at the O K ($1s\rightarrow2p$) edge. The horizontal scale is shifted by the energy of the final state core hole as measured by XPS. 

At the threshold the two absorption spectra are very similar, except for a broader leading edge in the O K case due to trivial intrinsic (shorter core hole lifetime) and extrinsic (experimental resolution) factors. However, the Ir N$_7$ XAS presents a clear pre-edge feature for the B surface. The strong difference in the pre-edge intensity is reproducible over several cleaves and is in some sense surprising given that XAS in total electron yield is generally considered rather bulk sensitive. The presence of an absorption channel below the threshold is the effect of a reorganization of the conduction states upon creation of the $4f$ core hole. The interaction with the final state core hole can shift the energy of the empty electronic state below the Fermi level and bring the XAS leading edge below the nominal threshold. As a consequence, the shape of the XAS spectrum may bear little resemblance to the density of the unoccupied $5d$ states \cite{toyozawa1,toyozawa2}. 

In a band insulator or semiconductor the addition of charge carriers fills defect states and induces a gradual shift of the bands, generally until a band edge is reached. Here, due to electron correlations, the addition of carriers results in a qualitatively different DOS, as sketched in the inset of Fig.~\ref{fig3} \cite{dosnote}. The states at the Fermi level in B provide new available absorption channels at lower energy than the first unoccupied J$_\mathrm{eff}$=1/2 level in A and yield the broad pre-edge peak which is vanishing in A \cite{xasnote}. In A the bulk DOS is hidden by the absence of the Na layer and ARPES does not detect the lowest energy state. This explains the poor agreement between the binding energy of the flat bands seen here and in Ref.~\onlinecite{comin}, and the $\sim$0.4 eV gap measured by optics. Notice that in this picture the lowest energy state is populated by electrons transferred from the Na $sp$ to the Ir 5$d$ states, and therefore ``inherits'' the Mott gap from the J$_\mathrm{eff}$=1/2 electrons. Further support for the presence of Ir $5d$ spectral weight in the vicinity of the Fermi level state comes from resonant photoemission data \cite{supplinfo}.

The available first principle calculations, in order to find consistency with the measured optical gap, tend to introduce a rather high value for $U$, typically 3-4 eV \cite{mazin2013,comin}. Probably for this reason, they also tend to overestimate correlations and yield bandwidths of 0.2-0.3 eV \cite{sohn,yunoki,mazin2012,comin}, at variance with what is observed here both for the dispersive band bottom of the lowest energy state, and for the $>$1 eV bandwidth of the hole-like states. However, in both Refs.~\onlinecite{sohn} and \onlinecite{comin} the onset of the optical conductivity is very gradual. It is entirely possible that the actual gap could be slightly smaller than reported and that calculations should be tuned to a lower value of $U$. Perhaps more importantly, our results show the decisive role of the Na electrons in shaping the magnetic and electronic structure of Na$_2$IrO$_3$. The ground state configuration may therefore have less $d^5$ character than generally assumed, with additional multiplet terms to take into account aside from the $d^5$ initial state J$_\mathrm{eff}$=1/2 doublet and J$_\mathrm{eff}$=3/2 quartet.\\

\acknowledgments

This research used resources of the Advanced Light Source, which is a DOE Office of Science User Facility under Contract No. DE-AC02-05CH11231. I.L.V. and A.L. acknowledge support from the U.S. Department of Energy, Office of Science, Office of Basic Energy Sciences, Materials Sciences and Engineering Division, under Contract No. DE-AC02-05CH11231 (Quantum materials KC2202). S.M. acknowledges support by the Swiss National Science Foundation under Grant No. P300P2-171221. S.U. acknowledges financial support from the Danish Council for Independent Research, Natural Sciences under the Sapere Aude program (Grant No. DFF-4090-00125). J.G.A acknowledges support towards the synthesis of materials from the Department of Energy Early Career program, Office of Basic Energy Sciences, Materials Sciences and Engineering Division, under Contract No. DE-AC02-05CH11231. N.P.B. and J.G.A. acknowledge support from the Gordon and Betty Moore FoundationÕs EPiQS Initiative through Grant GBMF4374

\bibliographystyle{apsrev}

\begin{thebibliography}{42}
\bibitem[{dos()}]{dosnote}
\bibinfo{note}{The presence of two different DOS on the surface of
  Na$_2$IrO$_3$ was already noticed by STM in Ref. 34. Here we reveal the
  connection to the surface termination and show that the Na-rich surface is
  the one with the narrower gap}.
\bibitem[{xas()}]{xasnote}
\bibinfo{note}{We believe that the small pre-edge seen there be due to deeper
  Na layers and to the larger probing depth of XAS with respect to ARPES}.
\end{thebibliography}

\end{document}